Salinity tolerance in plants: attempts to manipulate ion transport

Vadim Volkov

Faculty of Life Sciences and Computing, London Metropolitan University, 166-220 Holloway Road, London N7 8DB, UK

For correspondence: vadim.s.volkov@gmail.com and v.volkov@bio.gla.ac.uk

Phone: +44 (0) 20 7133 4043

Fax: +44 (0) 20 7133 4149

Ion transport is the major determining factor of salinity tolerance in plants. A simple scheme of a plant cell with ion fluxes provides basic understanding of ion transport and the corresponding changes of ion concentrations under salinity. The review describes in detail basic principles of ion transport for a plant cell, introduces set of transporters essential for sodium and potassium uptake and efflux, analyses driving forces of ion transport and compares ion fluxes measured by several techniques. Study of differences in ion transport between salt tolerant halophytes and salt-sensitive plants with an emphasis on transport of potassium and sodium via plasma membranes offers knowledge for increasing salinity tolerance. Effects of salt stress on ion transport properties of membranes show huge opportunities for manipulating ion transport. Several attempts to overexpress or knockout ion transporters for changing salinity tolerance are described. Future perspectives are questioned with more attention given to potential candidate ion channels and transporters for altered expression. The potential direction of increasing salinity tolerance by modifying ion channels and transporters is discussed and questioned. An alternative approach from synthetic biology is to modify the existing membrane transport proteins or create new ones with desired properties for transforming agricultural crops. The approach had not been widely used earlier and leads also to theoretical and pure scientific aspects of protein chemistry, structure-function relations of membrane proteins, systems biology and physiology of stress and ion homeostasis.

Salinity is among the most serious problems for modern agriculture with the estimated annual losses nowadays being over USD 12 billion (Pitman and Läuchli, 2002; Shabala, 2013). More than 20% of irrigated lands and up to 50% (Flowers, 1998) are affected by salinity, which essentially reduces the yield of agricultural crops since most of them are salt-sensitive glycophytes (Munns and Tester, 2008). After the collapse of Sumer civilization about four thousand years ago caused by improper agricultural techniques, which led to soil salinization and drop in the agricultural productivity in the area (Pitman and Läuchli, 2002) progress in agriculture and science rather allowed to set problems and pose questions than to get clear answers for dealing with salinity and growing salt-tolerant plants. The review provides basic details of ion transport via cell membranes, describes main transport systems for sodium and potassium, compares salt-sensitive glycophytes with salt-tolerant halophytes and considers genetically modified and artificially designed transport proteins for improving ion transport under conditions of salinization and salt treatment.

# Basic assumptions for studying ion transport in plant cells

Before pondering salinity tolerance and ways to change ion transport it's important to describe a typical plant cell and introduce basic parameters for understanding how ion fluxes change intracellular ion concentrations.

### *Cell size-volume-surface/volume ratio*

The sizes of cells within a plant may differ orders of magnitude, from small cells of root xylem parenchyma to huge epidermal cells in leaves. Cells of xylem parenchyma in roots of a model well-studied dicot plant *Arabidopsis* are less than 5 mkm in diameter with length often below 20-30 mkm (eg: Dolan et al., 1993; Ivanov, 1997; Kurup et al., 2005; Verbelen et al., 2006; Ivanov, Dubrovsky, 2013); the cells could be isolated and (after digesting the cell walls with cellulolytic enzymes) result in protoplasts of about 10 mkm in diameter compared to larger 20 mkm epidermal protoplasts from root elongation zone or 15-25 mkm protoplasts from root cortex parenchyma cells and the other root tissues (Demidchik, Tester, 2002; Demidchik et al., 2002; Volkov, Amtmann, 2007). The volumes would correspond to $4/3*\pi*R^3$ that is about 500 fL for 10 mkm protoplasts and about 8 pL for 25 mkm protoplasts (the protoplasts are used for different studies including electrophysiological recordings, so the values are calculated also for them).

Cells in leaf epidermis of several monocots are quite large. Barley epidermal leaf cells could be up to 2 mm long and about 25-30 mkm wide (with nearly square cross-section) reaching volume over 1000 pL (Volkov et al., 2007), though isolated from the cells protoplasts are usually smaller. Several studies involved isolated protoplasts from barley leaf epidermal cells and reported 60 mkm protoplasts (Dietz et al., 1992) corresponding to about 100 pL, 40 mkm (from 20 to 80 mkm) protoplasts (Karley, Leigh, Sanders, 2000) corresponding to about 30 pL (from to 4 pL to 250 pL), 25 pL protoplasts with large variations (over 10 times) in volume (Volkov et al., 2007) and 30 mkm protoplasts (Volkov et al., 2009) corresponding to about 14 pL. It is worth to mention 1) that larger cells might be lost or disintegrated during the isolation and also 2) that about 99% of large epidermal leaf protoplasts could be occupied by the vacuole (Winter, Robinson, Heldt, 1993).

Assuming volumes of usual plant cells within 500 fL to 1000 pL, the calculated surface to volume ratios will differ about 10 times for the cells (depending on the shape of the cells): from about 0.9 mkm$^{-1}$ to 0.16 mkm$^{-1}$ for oblong cells within plant tissues and from 0.6 mkm$^{-1}$ to 0.05 mkm$^{-1}$ for protoplasts of the corresponding volumes. It means that the ion fluxes via plasma membrane should be reverse proportional to surface to volume ratio to cause the same changes in ion concentrations for cells (Figure 1). The above mentioned smaller cells have about 0.9 mkm$^2$ of membrane surface per 1 mkm$^3$ of volume, while large cells just 0.16 mkm$^2$ per 1 mkm$^3$ and potentially need higher ion fluxes or larger time for the same concentration changes.

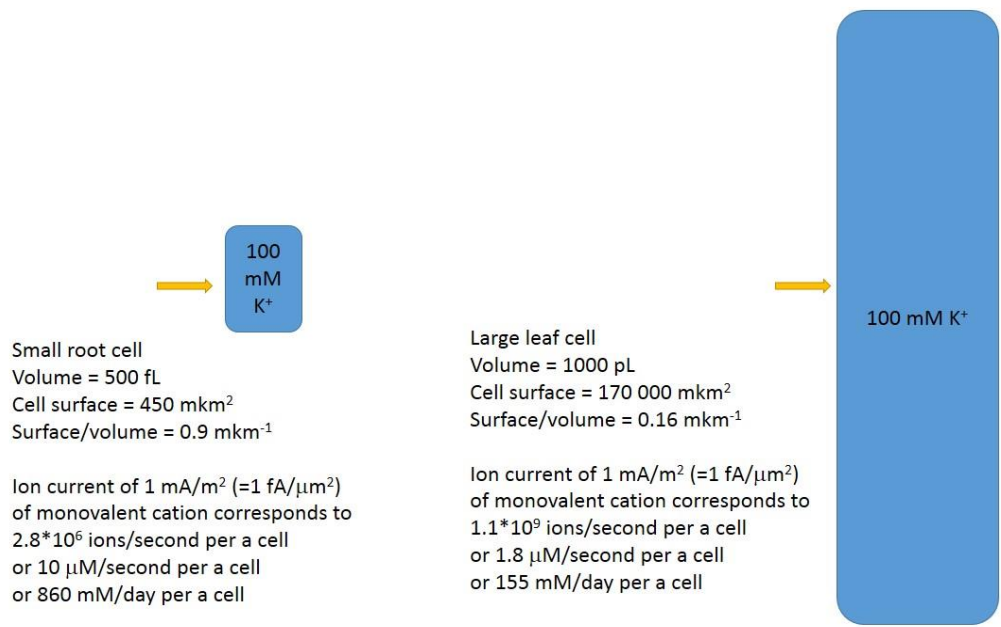

Figure 1. Simplified scheme of different plant cells and the ion fluxes for the cells. The same ion fluxes result in different changes of ion concentrations for cells of different sizes and surface/volume ratios. Ion fluxes with number of ions/second are directly recalculated to concentrations.

*Ion concentrations*

Typical potassium concentrations in cytoplasm of plant cells were measured independently by several methods (including ion-selective electrodes, fluorescent dyes and X-ray microanalysis) and range around 60-140 mM (Pitman, Läuchli, Stelzer, 1981; Hajibagheri et al., 1988; Hajibagheri, Flowers, 1989; Walker, Smith, Miller, 1995; Walker, Black, Miller, 1998; Korolev et al., 2000; Cuin et al., 2003; Halperin, Lynch, 2003; Shabala et al., 2006; Hammou et al., 2014), though concentrations above 200 mM were estimated by efflux analysis (reviewed in Britto, Kronzucker, 2008) and as low as 12 mM and even lower concentrations were measured by ion-selective electrodes in root cells of potassium-deprived *Arabidopsis* plants (Armengaud et al., 2009). Higher potassium concentrations of 200-350 mM measured in cell sap by X-ray microanalysis or capillary electrophoresis are rather attributed to vacuolar compartment under sufficient potassium supply (Malone, Leigh, Tomos, 1991; Fricke, Leigh, Tomos, 1994; Bazzanella et al., 1998; Volkov et al., 2004). It is worth to mention that in animal cells potassium seems to be among regulators of apoptotic enzymes activating them at 50 mM and lower $K^+$ concentrations (Hughes, Cidlowski, 1999).

Sodium cytoplasmic concentrations of plant cells are usually low reaching about 20-50 mM after several days of NaCl treatment (Carden, Diamond, Miller, 2001; Carden et al., 2003; Halperin, Lynch, 2003). Higher sodium concentrations had been also measured depending on duration of salt stress, external sodium and the other ions concentrations and plant species (cytoplasmic sodium concentrations over 200 mM were reported in salt-tolerant halophytes) (Hajibagheri, Flowers, 1989; Halperin, Lynch, 2003; Flowers, Colmer, 2008; Kronzucker, Britto, 2011).

In halotolerant alga *Dunaliella salina*, which is a good unicellular eukaryotic model for studying and understanding salinity tolerance and ion transport within the range of 0.05 -5.5 M NaCl (e.g. Katz, Avron, 1985), cytoplasmic sodium concentrations about 90 mM (88 ± 28 mM) were reported using $^{23}$Na-NMR spectroscopy (Bental, Degani, Avron, 1988). It's interesting that $Na^+$ concentrations were nearly the same (within the error of a few measurements) in the algal cells adapted to a wide range of external $Na^+$, from 0.1 M to 4 M (Bental, Degani, Avron, 1988). Similar or even lower sodium concentrations below 100 mM were measured also by the other methods for the alga under 0.5-4 M or 1-4 M sodium treatment (Katz, Avron, 1985; Pick, Karni, Avron, 1986). The small alga *Dunaliella salina* has length about 10-11 μm, width of 6 μm and volume around 200 fL (Masi, Melis, 1997) or even smaller dimensions with volume around 90-100 fL then (Katz, Avron, 1985). It probably possess specific transport system to exclude $Na^+$, similar to $Na^+$-ATPase expected to function in the plasma membrane of the marine unicellular alga *Platymonas viridis* (Balnokin, Popova, 1994), alga *Dunaliella maritima* (Popova et al., 2005) and several other marine algae (reviewed in: Balnokin, 1993; Gimmler, 2000). Genome and transcriptome sequencing of the algae, isolation and heterologous characterisation of the corresponding genes and proteins will provide more information about their transport systems (see below).

However, high sodium concentrations over 100 mM often have inhibiting effect on protein synthesis at least in salt-sensitive glycophytes (Hall, Flowers, 1973; Wyn Jones, Pollard, 1983; Flowers, Dalmond, 1992). Sodium is also 1) competing with potassium for allosteric sites of enzymes, for example, and 2) interacting with ion channels (for example, sodium ions change the gating of potassium outward rectifying currents in root protoplasts of halophyte plant *Thellungiella*, which are most likely carried by Shaker type potassium channels: Volkov, Amtmann, 2006). Moreover at the cellular level salt stress induces apoptosis (Katsuhara, Kawasaki, 1996; Huh et al., 2002; shortly reviewed in Shabala, 2009; Demidchik et al., 2010).

Much higher sodium concentrations could be tolerated in vacuoles, one of the functions of the organelle is to sequester and isolate sodium. Concentrations of sodium in vacuoles may exceed 0.5-1 M being up to ten times over the cytoplasmic sodium concentrations (eg Flowers, Troke, Yeo, 1977; Zhao et al., 2005; Flowers, Colmer, 2008) at the expenses of activity of specific ion-transport systems (reviewed in: Martinoia et al., 2012). Under salt treatment of 2 M NaCl for 85 days shoot tissue concentrations of sodium in halophytes *Tecticornia* were about 2 M, so presumably vacuolar $Na^+$ concentrations could be over 2 M in the halophytes (English, Colmer, 2013).

The reasons for sodium competing with potassium are that potassium and sodium ions have 1) the same electric charge, $1.6*10^{-19}$ coulombs, 2) similar cation radii in non-hydrated, about 0.1 nm = 1.0 Å for sodium cation and 0.14 nm = 1.4 Å for potassium (diameter being nearly 2-3% of cell membrane thickness) and 3) hydrated forms, about 3.6 Å for sodium and 3.2 Å for potassium ions (Nightingale, 1959; Collins, 1997; Mähler, Persson, 2012) and, hence, 4) similar surface electric charge densities, which differ about 2 times (twice higher for non-hydrated sodium according to charge and diameter of the ion). Therefore interactions of $Na^+$ and $K^+$ with amino acids of protein surfaces, active centers of enzymes, pockets of allosteric regulation or binding of proteins, selectivity filters of proteinaceous ion channels are similar and often differ only several times in selectivity of the interactions. The selectivity depends on the nature and number of interacting amino acids and their spatial location. It is interesting to mention that molecular dynamics simulations together with conductivity measurements for several proteins and oligopeptides demonstrated higher (up to 5 times) affinity of sodium over potassium to the protein surfaces (especially with numerous carboxyl groups) (Vrbka et al., 2006). The phenomenon could explain higher destabilising effect of sodium over potassium on proteins ("salting them out"), which was initially discovered with white proteinaceous part of hen's eggs by Hofmeister in 1888 (Hofmeister, 1888; Kunz, Henle, Ninham, 2004). Higher destabilising effect of sodium on proteins could be the reason why potassium and not sodium is chosen and naturally selected for being the major intracellular monovalent cation, pumped into cells while pumping out sodium cations (Collins, 1997) though sea and ocean water contains more than 40 times higher concentration of sodium. Under salt stress, for plants it's important to keep higher $K^+/Na^+$ ratio (Maathuis, Amtmann, 1999). It's essential, however, to mention that some proteins (due to specific amino acid composition or structural peculiarities) and processes from halophytes are able to withstand higher sodium concentrations without loss of activity (eg. Flowers, Dalmond, 1992; Premkumar et al., 2004); it seems to be the secondary evolutionary adaptation since halophytes presumably emerged many times independently (Bennett, Flowers, Bromham, 2013). It's also interesting that cell wall proteins of studied halophytes and also glycophytes did not change their activity within wide range of sodium concentrations, often from 0 to over 0.5-1 M (Thiyagarajah, Fry, Yeo, 1996).

### *Driving forces and pathways for ion transport to cells*

Transport of ions is driven by physico-chemical forces including differences of ion concentrations (or to be more precise, activities of ions) and differences in electric potential at the sides of membranes.

Membrane potential of plant cells is routinely measured and monitored by microelectrodes with tiny sharp tips around 0.1 mkm in diameter after impalement of a plant cell of interest (eg. described in: Blatt, 1991). Recently developed voltage-sensitive fluorescent proteins and dyes (reviewed in: Mutoh, Akemann, Knöpfel, 2012) could be also used for at least indications of membrane potential in cell tissues and populations of cells (Matzke, Matzke, 2013). Membrane potentials below -70 mV and above -200 - -220 mV are recorded by microelectrodes though values around -300 mV were also reported; lower (more negative) membrane potential values are often measured in root cells compared to leaf ones apart from leaf guard cells (Higinbotham, 1973; L'Roy, Hendrix, 1980; Blatt, 1987; Walker, Smith, Miller, 1995; Walker, Black, Miller, 1998; Carden, Diamond, Miller, 2001; Shabala, Lew, 2002; Carden et al., 2003; Fricke et al., 2006; Murthy, Tester, 2006; Shabala et al., 2006; Volkov, Amtmann, 2006; Armengaud et al., 2009; Hammou et al., 2014). Vacuolar membrane potential is the same or 10-40 mV above the values for cytoplasm with pH about 2 or over units lower, about 5.0-6.1 or less in vacuoles compared to 7.0-7.7 in cytoplasm (eg. Walker, Smith, Miller, 1995; Carden et al., 2003; Cuin et al., 2003; Martinoia et al., 2012) (Figure 2).

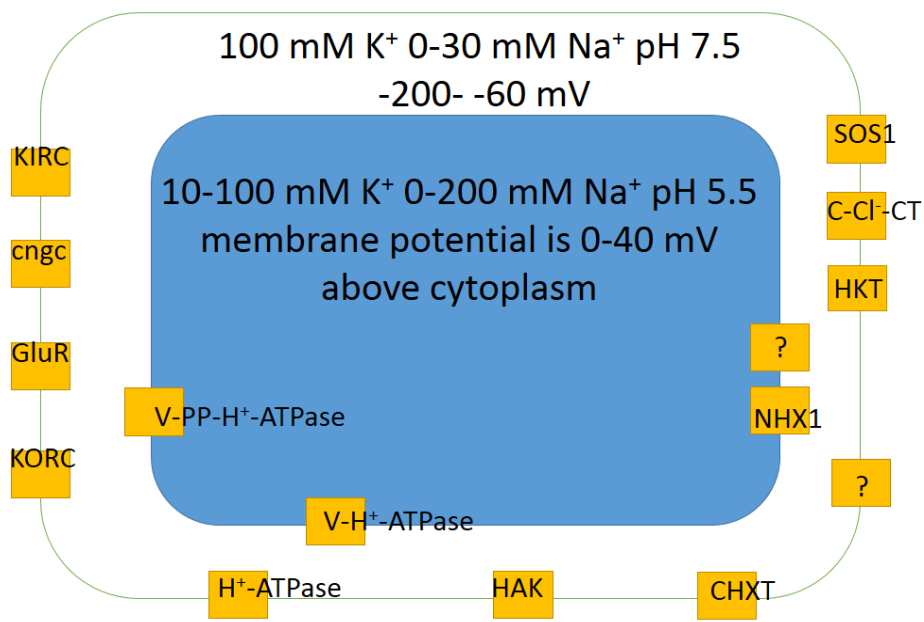

Figure 2. Basic scheme of membrane potentials, potassium $K^+$ and sodium $Na^+$ concentrations and pH values in a generalised plant cell together with main ion transport systems ensuring potassium and sodium transport according to electrochemical forces. Different cell types usually have less transporters, though specialised for more determined ion transport functions. The concentrations and membrane potentials are rather indicative and change depending on conditions of mineral nutrition and are not the same for different cell types (see text and references for more details). KIRC are inward rectifying potassium channels (eg. Hirsch et al., 1998); KORC are outward rectifying potassium channels; GluR

are glutamate receptors; cngc are cyclic nucleotide gated ion channels; HAK is high affinity potassium transporter; CHXT are cation $H^+$ exchange transporters (eg. Evans et al., 2012); HKT are high $K^+$ affinity transporters; C-$Cl^-$-CT are cation chloride contrasnporters; SOS1 is well studied sodium-proton antiporter; $H^+$-ATPase is proton pump of plasma membrane; V-$H^+$-ATPase is vacuolar proton pump; V-PP-$H^+$-ATPase is vacuolar pyrophosphatase, another vacuolar ptoton pump; NHX1 is vacuolar sodium (cation)/proton antiporter. For more details and description see text.

Thermodynamics of ion transport is basically relatively simple (without peculiarities) and described by several equations. Nernst equation applied for selectively permeable membrane links ion concentrations at the sides of membrane with the electric potential via the membrane under equilibrium conditions, when net flux of ions via the membrane is absent:

$E_S = E_1 - E_2 = R*T/(Z_S*F) * \ln([S_2]/[S_1])$

(Hille, 2001). Here E is electric potential, R is universal gas constant equal to 8.31 J/(K*mole), T is temperature in K, $Z_S$ is the charge of ion S, F is Faraday constant equal to 96 500 s*A / mole, $[S_1]$ and $[S_2]$ are concentrations of ion S at the sides of the membrane. Basically, the diffusion of ion S due to different concentrations is equilibrated by the electric potential, which is about ± 60 mV (slightly depending on temperature) per ten-fold difference in concentrations with sign determined by the ion charge. For example, for potassium with typical 100 mM in cytoplasm of epidermal root cell and low 100 μM in soil solution the membrane potential for inward flux (uptake) of $K^+$ should be below -180 mV to satisfy the electrochemically downhill transport of $K^+$ ions. Lower concentrations of potassium outside the cells may require co-transport of $K^+$ with the other ions (e.g. with $H^+$) to ensure the energy demands for the transport (see below).

For several ion species with specific permeabilities via the membrane a more complicated Goldman-Hodgkin-Katz voltage equation is applicable; it takes into account permeabilities of the ions. For sodium, potassium and chloride (obviously more ions to be considered and more components should be added) the equation will be:

$E_{membrane} = R*T/F * \ln((P_{Na^+}[Na^+]_{out} + P_{K^+}[K^+]_{out} + P_{Cl^-}[Cl^-]_{in})/(P_{Na^+}[Na^+]_{in} + P_{K^+}[K^+]_{in} + P_{Cl^-}[Cl^-]_{out}))$,

where $E_{membrane}$ is membrane potential via the membrane (or $E_{reversal}$ with zero net ion current via the membrane), P are permeabilities of the corresponding ions and [] stands for concentrations of the ions (Hille, 2001). Usually potassium permeability is dominating and membrane potential is close to the $E_{reversal}$ of $K^+$, though may change and depend on the cell type. Plasma membrane proton pump $H^+$-ATPase (reviewed e.g. in: Palmgren, 2001) usually shifts membrane potential to more negative values compared to the calculated $E_{reversal}$ for $K^+$, while $K^+$ or $Na^+$ pumps are not known for plants.

Transport of most ions including $Na^+$ and $K^+$ in plants occurs passively (following the electrochemical forces) via ion-selective proteinaceous pores of ion channels, which can change their conformation from open to close and vice versa (so called "gated") under applied voltages or after binding ligands and regulators. Another pathway is via proteinaceous transporters with slower transport rates. Ion transport via ion channels is electrogenic since ions carry electric charge, while transporters realise electrogenic or non-electrogenic transport, transporting one ion or co-transporting/antiporting several charged ions in one or opposite directions, correspondingly.

Co-transport of several ions or even small molecules may add extra energy for transport. For example, HAK transporters presumably co-transport $K^+$ together with $H^+$ (Banuelos et al., 1995; Rodriguez-Navarro, 2000; Grabov, 2007), which gives several orders of concentration differences extra due to transport of $H^+$ according to electric charge. Membrane potential of -180 mV potentially allows potassium uptake, when co-transported with $H^+$ with e.g. stoichiometry 1:1 under similar external pH to pH of cytoplasm, against $10^6$ difference in $K^+$ concentrations (e.g. Rodriguez-Navarro, 2000). Higher concentrative capacity could be achieved using also pH differences or higher number of protons per $K^+$; cytoplasmic pH is about 7.5 and external low pH of e.g. 4 will add over three orders more. The surprising example is described for yeast *Schwanniomyces occidentalis*, which was reported to deplete external potassium to 0.03 µM, so presumably taking up $K^+$ against 3 000 000 difference in concentrations (assuming over 100 mM of cytoplasmic $K^+$) due to HAK transporters (Banuelos et al., 1995). Much higher concentrations, around 80 µM, arising from $K^+$ contamination from agar and the other chemicals (Armengaud et al., 2009; Kellermeier et al., 2014) (though contamination from agar was estimated at 1-3 µM: Kellermeier et al., 2014) resulted in symptoms of severe potassium deficiency in *Arabidopsis* and essentially changed transcription profile in the roots and shoots of the plants (Armengaud, Breitling, Amtmann, 2004), so more detailed examination with special attention to transport systems of different species is required. Basically a set of transporters and ion channels is specific for cell types and organisms, includes tens and more distinct characterised so far proteins, which often form heteromers with variable properties and regulation (Figure 2). Detailed analysis of genome sequences of salt-sensitive model plant *Arabidopsis* revealed that about 5% of about 25,000 genes of the plant potentially encode membrane transport proteins; the genes of the about 880 proteins are classified in 46 unique groups, while genes of cation channels/transporters predict for coding over 150 proteins (Mäser et al., 2001). Special databases include information about transport proteins, e.g. plant membrane transport database http://aramemnon.botanik.uni-koeln.de/ ; http://www.yeastgenome.org/ is a useful source of information for yeast proteins including yeast membrane transport proteins.

### *Ion channels vs ion transporters: more about pathways of ion transport to cells*

One of disputable questions of ion transport is linked to ion channels and transporters and their role in ion transport. It is commonly assumed that ion channels in an open state/conformation allow passage of over $10^7$-$10^8$ ions per second via a selective pore formed within a protein molecule. The diameter of the pore is determined by the molecular structure of ion channel, from 12 Å for potassium channel KcsA with narrow part of 4 Å in diameter (eg. Doyle et al., 1998; Jiang et al., 2002; MacKinnon, 2004) to 15 Å and even 28 Å diameters of pores for the general bacterial porins with low selectivity and permitted passage for small hydrophilic molecules (about 6 Å pores for the highly selective porins) (eg.: Galdiero et al., 2012). The diameter of the pore and amino acids lining it essentially determine the ion selectivity of ion channel and potential number of passing ions per a unit of time. The selectivity could be, for example, over 1000 for $K^+$ over $Na^+$ in potassium selective ion channels or over 10 for $Na^+$ over $K^+$ in sodium selective channels due to special selectivity filters with conserved amino acids for specific channel types; often amino acid sequence glycine-tyrosine-glycine GYG indicates selectivity for $K^+$, introducing mutations into the pore to change the amino acids converted potassium selective ion channels to nonselective ones (eg. Heginbotham, Abramson, MacKinnon, 1992). The interactions of ions with the protein molecule of ion channel are not well understood yet and probably involve non-electrostatic ion-ion interactions, van der Waals forces, interaction with water molecules and numerous other interactions, so several methods of modelling and simulations of molecular dynamics are applied within at least the last 30 years; the sharp increase in computing power allowed to include the lipid environment of membranes, pH and the known biochemical factors and regulators to the models (eg. reviewed comprehensively with over 600 references in: Maffeo et al., 2012).

Direct measurements are the basement for investigating ion fluxes via ion channels; they provide information about permeating ions, number of the ions per a second, selectivity and potential complex transitions of protein molecules of ion channels during the transport processes. Indeed, a small current of 1 pA corresponds to

$10^{-12}$ A/($1.6*10^{-19}$ C) ≈ $6*10^6$ ions/second ($1.6*10^{-19}$ C is elementary charge, a charge of monovalent cation), while most ion channels demonstrate much larger electric currents with complex voltage-dependent patterns of open-closed states (Figure 3).

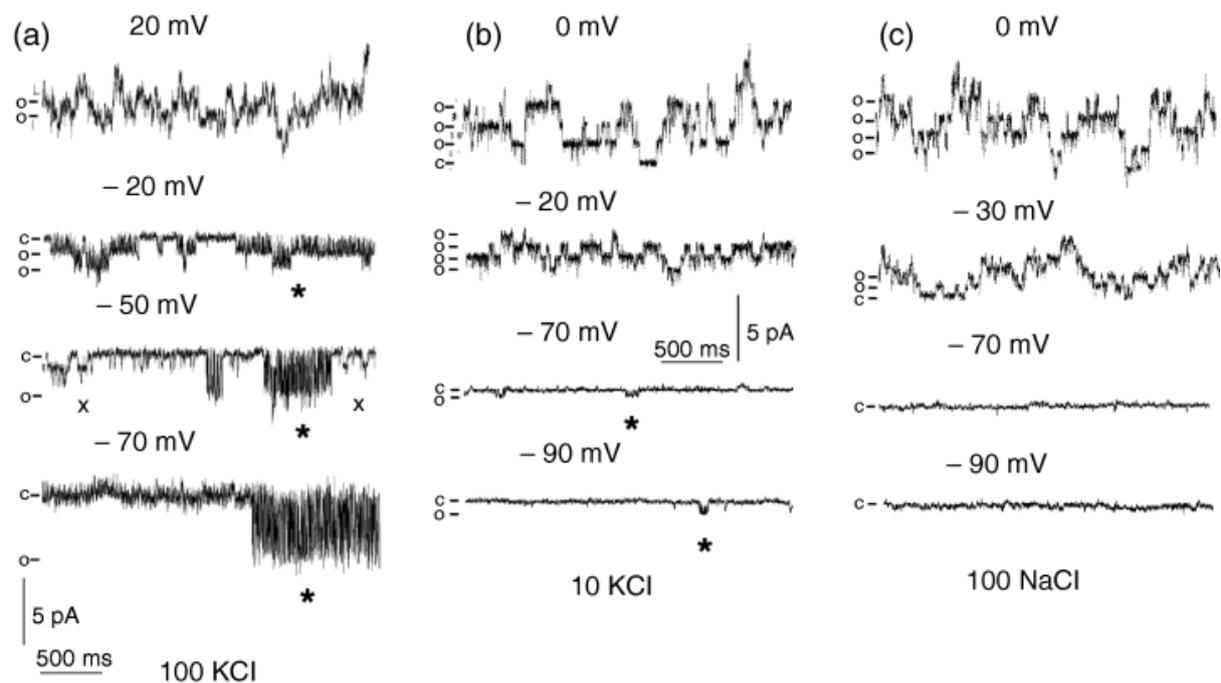

Figure 3. Single-channel recordings in outside-out patches of *T. halophila* root protoplasts. The pipette solution was 100 mM KCl. The bath solution contained 100 mM KCl (a), 10 mM KCl (b) or 100 mM NaCl (c). c, current level with no open channels; o, current levels of single-channel openings. Spiky openings of outward-rectifying channels allowing inward current are indicated with asterisks. Openings of a second type of channel are indicated with crosses. Reproduced from (Volkov, Amtmann, 2006) with the permission from the publisher John Wiley and Sons.

Transporters are a sort of "enzymes" with conformational changes of a protein molecule required for a complete transport cycle of ions (turnover of the transporter) (eg. Gadsby, 2009). The lower estimate for turnover of transporters is the activity of ion pumps. Plant plasma membrane $H^+$-ATPase pumps about 100 ions per a second (Sze, Li, Palmgren, 1999). The values are comparable to yeast plasma membrane ATPase Pma1: one molecule of Pma1 can transport 20-100 $H^+$ per second (Serrano, 1988; and also estimates from specific ATP hydrolysis activities from e.g. Perlin et al., 1989). The value is also similar to animal $Na^+/K^+$-ATPase with a turnover of 160 (Skou, 1998). Similar or even lower turnover rates, from 3 to 60, were shown for sodium/glucose cotransporter (Longpré, Lapointe, 2011 and references therein), while turnover about 500 was estimated for sucrose/$H^+$ co-transporter from maize ZmSUT1 (Carpaneto et al., 2010).

The highest possible rate for activity of ion transporters could be assessed from protein structure studies and frequency of conformational changes with estimated upper limit of $10^6$ (Chakrapani, Auerbach, 2005), which seems to be an overestimated value. A more realistic value for the fast transporters/pumps is around 10,000 ions/second, when they are mutated and have accelerated turnover (Gadsby, 2009). A higher value of $10^5$ ions per a second was reported for $Cl^-/H^+$ antiporter ClC-5, which is rather a unique type of transporter similar to ClC channels (Zdebik et al, 2008).

Unfortunately, ion fluxes via a single transporter (order of several fA or much lower) are below the resolution for measurements: the most sensitive so far electrophysiological method of patch clamp can hardly provide stable recordings below 0.1 pA due to technical limitations of equipment and the corresponding unavoidable noise. However, potentially the ion currents via at least hundreds and rather thousands or millions of electrogenic (transferring electric charge during a transport cycle) ion transporters could be measured under specific conditions. A report about unitary conductance of $Cl^-/H^+$ antiporter ClC-5 is an exception, the conductance of 0.45 pS (ion current about 63 fA at 140 mV) for the transporter was determined from noise analysis of recordings with hundreds/thousands of transporters (Zdebik et al., 2008). The unusual very high conductance of a typical transporter HKT is reported for rice OsHKT2;4; the transporter was expressed in *Xenopus* oocytes and single channel traces with a few pA open/closed states were recorded (Lan et al., 2010). However, the recordings should be rather attributed to a behaviour of an ion channel, but not a transporter and posed questions about the structure and properties of OsHKT2;4 (Lan et al., 2010); further on the results with expression of OsHKT2;4 were not confirmed and the recordings were considered to be endogenous currents of *Xenopus* oocytes (Sassi et al., 2012).

Often transporters are heterologously expressed in large *Xenopus* oocytes, where detectable ion currents or fluxes of radioactive tracers are reported after the expression at the background of usually small endogenous electric currents of the oocytes. The successful expression in *Xenopus* oocytes was reported for several HKT transporters and cation-chloride cotransporters, attempts to record activity of HAK, SOS1, or Nhx transporters were less fruitful so far (Rodriguez-Navarro, 2000; Liu et al., 2001; Colmenero-Flores et al., 2007; Jabnoune et al., 2009; Rodríguez-Rosales, 2009; Eduardo Blumwald, personal communication). A typical surface area of an oocyte with diameter around 1 mm (mature *Xenopus* oocytes used for heterologous expression are quite even and have the same diameter of 1 mm) will be $4*\pi*500*500$ mkm$^2$ ≈ $4*3.14*250000$ ≈ 3,100,000 mkm$^2$. Assuming the very high recorded so far values of about -10 mkA at -150 mV for heterologous expression of rice transporter HKT in *Xenopus* oocytes (Jabnoune et al., 2009) being reasonable and not due to incorrect folding/partial proteolysis/interaction with endogenous transport systems of *Xenopus* oocytes, we can recalculate the current per a unit of oocyte surface and compare with recordings from plant protoplasts:

-10 mkA/3,100,000 mkm$^2$ ≈ 3 pA/mkm$^2$ = 3,000 mA/m$^2$ or about 1 nA per a small root protoplast with diameter around 10 mkm. The values are very high and comparable or even much higher (see below) than the ones recorded from activity of ion channels using patch clamp.

Another theoretical estimate is useful for assessing activity of ion transporters expressed in *Xenopus* oocytes. Assuming the very high turnover for the transporter being around $10^4$ and the very high expression of about 10,000 transporters/mkm$^2$ (means a transporter per 10*10 nm$^2$, nearly maximal due to the physical and steric limitations) we get the possible presumed current per an oocyte of *Xenopus*:

$10^4$ ions per transporter per second*3,100,000 mkm$^2$ *10,000 transporters/mkm$^2$ = $3,1*10^{14}$ ions/second, then taking elementary charge of monovalent ion being $1.6*10^{-19}$ C, the estimate gives $1.6*10^{-19}$ C*$3,1*10^{14}$ ions/second ≈ 50 mkA per an oocyte. An excessive order of magnitude could be reasonably explained by a lower level of expression and lower transport rate of a transporter, so gives a reasonable agreement with experimental data and leads to several conclusions.

1) Transporters can provide sufficient ion currents for registered ion transport under conditions of salinity.

2) Ion channels are not the only essential pathway for ion transport under salinity.

3) Balance between relative share of ion transport by ion channels or by ion transporters depends on abundances of the corresponding proteins (ion channels or transporters), their regulation and the other factors (composition of ion solutions, membrane potential etc.). Potentially transporters may ensure fine-tuning of ion transport, while ion channels provide large ion fluxes when required.

The estimates provide a potential positive answer for a question whether total ion current via thousands/millions of electrogenic transporters could be measured and characterised in plant protoplasts using patch clamp in whole-cell configuration (recording sum of all ion currents via the whole membrane). Well-studied so called nonselective cation channels with low conductance carry small instantaneous currents and potentially the total current via numerous transporters could be of the same range. However, the exact mechanisms of ion transport by different ion transporters are not always clear and make the further predictions complicated. For example, HKT transporters could be similar to ion channels forming a specific ion-selective pore with properties distinct from the pore of ion channels (the pore of HKT has an extra amino acid constriction making an additional energy barrier for ion transport) (eg reviewed for HKT transporters in: Yamaguchi, Hamamoto, Uozumi, 2013; Benito et al., 2014) according to structure analysis (Cao et al., 2011). It means that transporters can electrically (from the point of ion electric current-voltage IV curve) behave in a way similar to ion channels and the reversal potential of ion current mediated by transporters can shift following ion concentrations inside and outside the cell. The situation is observed in *Xenopus* oocytes expressing different HKT transporters (Jabnoune et al., 2009; de Almeida, 2014), where rectification and Nernstian shift (according to Nernst equation for ion concentrations and voltages) in reversal potential were observed. Ion current via ion-selective ion channels is described by Goldmann-Hodgkin-Katz equations based on assumption of independent passage of ions via channel pore (or constant electric field along the diffusion zone) (Hille,

2001), additional charges of lipid bilayer and the surface of ion channel can modify the ideal curves (Alcaraz et al., 2004). However, similar to Goldmann-Hodgkin-Katz curves were recorded for HKT transporters expressed in *Xenopus* oocytes (Jabnoune et al., 2009; de Almeida, 2014).

Recent indications provide evidence that sodium currents via AtHKT1;1 transporters could be measured in *Arabidopsis* root stelar protoplasts (Møller et al, 2009; Xu et al, 2011). Patch clamp experiments demonstrated that protoplasts isolated from mature root stele of mutant plants overexpressing AtHKT1;1 in the tissue had higher sodium currents than control protoplasts (Møller et al, 2009). Control protoplasts in the study (line J2731* was taken) did not show any dependence of inward ion currents (about -35 mA/m$^2$ at -120 mV) on external Na$^+$ concentrations (0, 10 and 25 mM). However, protoplasts overexpressing AtHKT1;1 had similar to control currents in external 0 mM Na$^+$ (internal Na$^+$ was kept 30 mM in all the experiments), about 30 mA/m$^2$ lower (more negative) currents in 10 mM external Na$^+$ and about 50 mA/m$^2$ lower currents in 25 mM external Na$^+$ at -120 mV, when compared to control protoplasts. The currents in AtHKT1;1 overexpressing protoplasts also shifted the reversal potential according to external Na$^+$ concentrations, the shift confirmed Na$^+$ selectivity. Further study with another line E2586 (Xu et al, 2011) compared Na$^+$ and K$^+$ currents in root stellar cell protoplasts from wild type (control) and *athtk1;1-4* mutant plants lacking AtHKT1;1. Potassium currents were similar for the protoplasts of control E2586 and mutant plants, about -50 pA at -120 mV; the currents were recorded in 5 mM internal/50 mM external K$^+$ and corresponded to about -30 mA/m$^2$ at -120 mV, the reversal potential shifted according to the expected selectivity for potassium ions (about +50 mV). The protoplasts under investigation from root stele had an average diameter of 23 μm (Xu et al, 2011). Sodium currents were about -50 pA at -120 mV in control E2586 protoplasts at 50 mM internal/50 mM external Na$^+$ compared to about -10 pA in *athtk1;1-4* mutant ones and demonstrated Nernstian shift of reversal potential to +50 mV in 5 mM internal/50 mM external Na$^+$ with currents about -20 pA at -80 mV (approximating from the IV curves to about -30 pA at -120 mV) (Xu et al, 2011). The results pose numerous questions for future study. It had been clearly demonstrated that nonselective cation currents in *Arabidopsis* root protoplasts of 15- to 25-μm diameter (e.g. Demidchik, Tester, 2002 and see below) are slightly (1.5 times) more selective for potassium over sodium, so predictably *athtk1;1-4* mutant protoplasts would have 2-4 times higher sodium currents than measured in (Xu et al, 2011) due to expected nonselective cation currents.

The nonselective cation currents are studied well for root protoplasts, especially in *Arabidopsis* and carried by cyclic nucleotide gated channels (about 20 genes for *Arabidopsis*) and glutamate receptors (about 20 genes for *Arabidopsis*) (see below). One of the possible explanations for the paradox (apart from potential changes in ion transport due to mutant phenotype, changes in regulation of ion transport etc.) is that nonselective cation currents in root stelar protoplasts of *Arabidopsis* are highly selective for potassium over sodium; the selectivity was shown for root protoplasts of *Thellungiella* (Volkov, Amtmann, 2006), salt-tolerant relative of *Arabidopsis* (Bressan et al., 2001; Inan et al., 2004; Amtmann, 2009). It is important to understand specific tissue and cell type expression of genes and proteins for nonselective cation channels and HKT transporters for characterising their role in making the total ion currents. So far, nearly all the studies were done with nonselective cation currents, which sufficiently and completely explained the obtained experimental results.

Special modelling for different proteinaceous pores of ion channels will help to understand better the peculiarities of ion currents via ion channels and HKT-like transporters. Pharmacological analysis and profiling of ion currents is also essential together with the further use of mutants (knocking out or overexpressing specific transporter or channel of interest) and heterologous expression of the genes of interest in cell culture or in *Xenopus* oocytes. Nonselective ion currents are well characterised electrophysiologically and pharmacologically, especially for root protoplasts (White, 1993; White, Lemtiri-Chlieh, 1995; Roberts, Tester, 1997; Tyerman et al., 1997; Tyerman, Skerrett, 1999; Demidchik, Tester, 2002; reviewed in: Demidchik, Davenport, Tester, 2002; Volkov, Amtmann, 2006; reviewed in: Demidchik, Maathuis, 2007), while less is known for transporters. However, the research is progressing and recently quinine (500 $\mu$M) was shown to have slight inhibiting effect on ion currents induced by HKT1;4 transporters from durum wheat after heterologous expression in *Xenopus* oocytes; $Zn^{2+}$, $La^{3+}$, $Gd^{3+}$, or amiloride had no effect (Ben Amar et al., 2014). For cation-chloride co-transporter from *Arabidopsis thalina*, which was expressed in *Xenopus* oocytes and presumably transported $Na^+$:$K^+$:$2Cl^-$ 100 $\mu$M bumetanide had an inhibiting effect on uptake of radioactive ions similar to analogous animal co-transporters (Colmenero-Flores et al., 2007). Amiloride is shown to inhibit Nhx1 vacuolar $Na^+/H^+$ antiporter (Barkla et al., 1990; Darley et al., 2000). Experiments with heterologous expression of rice HKT transporter OsHKT2;4 in *Xenopus* oocytes demonstrated channel-like behaviour with single channel traces and inhibition by $Ba^{2+}$, $La^{3+}$ and $Gd^{3+}$ (Lan et al., 2010), however, the properties are not typical for a transporter and further on the results were not confirmed and attributed to endogenous currents of the expression system (Sassi et al., 2012).

A further complication for understanding the existing pathways of membrane ion transport comes from electroporation experiments (eg. Pakhomov et al., 2009; Wegner et al., 2011; Wegner, 2013; Wegner et al., 2013; Wegner, 2014). The nanopores with diameter about 1.0 nm (10 Å) formed in lipid membrane bilayer of several animal cell lines including GH3 and CHO-K1 after 600 nanosecond voltage pulses of 2.4 kV/cm and over; the pores persisted for minutes and provided additional inward rectifying and voltage-sensitive ion currents similar to ion currents in untreated cells (Pakhomov et al., 2009). The nanopores disappeared in minutes or broke into highly conductive, non-rectifying pores (Pakhomov et al., 2009). Similar nanopores with diameter about 1.8 nm (18 Å) were discovered in plant protoplasts derived from tobacco cell line BY-2 1) after 10 ms voltages pulses of 1 V and lower to -350-+250 mV and 2) also during patch clamp experiments after applied holding voltages above +200-250 mV or below -300 - -250 mV (Wegner et al., 2011; Wegner, 2013; Wegner et al., 2013). The electric conductance of the membrane rose 10 times and over after the treatment though high selectivity for $K^+$ over gluconate retained and was also reflected in reversal potential of current-voltage IV curves (Wegner et al., 2011). The voltage-induced nanopores demonstrated selectivity for cations (including even $TEA^+$) over anions ($Cl^-$) and slight selectivity for different cations ($Ca^{2+}$ and $Li^+$ were the most permeable); certain similarity with behaviour of nonselective cation channels was found (Wegner et al., 2011; Wegner, 2013; Wegner et al., 2013). Obviously, the results should be taken into consideration when analysing and planning experiments to avoid potential artefacts and obtain the realistic pattern of ion pathways into cells under determined physiological conditions (eg. voltages below -300 mV were recorded for plant cells). Moreover, 1) heterologous expression of specific ion channels, 2) using knockout and overexpressing mutants, 3) revealing detailed properties, pharmacological profiles and peculiarities of ion fluxes is necessary to exclude misinterpretation.

*Comparison of ion fluxes measured by different methods*

A few direct and indirect methods are used to determine ion fluxes to plant cells under control conditions and salt stress. They include 1) estimates based on kinetic measurements of ion concentrations (both in plants and in the nutrient solution), 2) electrophysiological methods, 3) technique of measuring fluxes using vibrating ion-selective electrodes (MIFE: microelectrode ion flux estimate/measurements), 4) measurements of unidirectional ion fluxes (e.g. radioactive isotope $^{22}Na^+$ for sodium, while $Rb^+$ is often used to imitate $K^+$ fluxes) and 5) several other methods using ion-selective fluorescent and non-fluorescent indicators, potentially NMR spectroscopy, ion-conductance scanning microscopy etc. It is reasonable to compare the values obtained by different methods for salt-tolerant and salt-sensitive plants.

Estimates from ion concentrations are the sum of influx and efflux of ions from plant cells, so the measurements provide essential physiological information about averaged net changes and fluxes. From the other point, since often the existing concentrations are already high, it takes large periods of time, at least hours and sometimes days (sufficient to determine the kinetic changes in ion concentrations) to obtain kinetic curves, usually without detailed peculiarities and lacking high temporal resolution of minutes. For example, after 25 hours of 100 mM NaCl treatment sodium contents in roots of glycophyte *Arabidopsis thaliana* rose from 0.2% to 2 % of dry weight (DW) of the roots (0.087 to 0.87 mM $Na^+$ per g of DW), while in halophyte *Thellungiella halophila* sodium in roots increased from 0.15% to 1.2% (0.065 to 0.52 mM $Na^+$ per g of DW); sodium contents in shoots of *Arabidopsis* changed from 0.3% to 1.7% (0.13 to 0.74 mM $Na^+$ per g of DW) and in shoots of *Thellungiella* the rise was from 0.8% to 1.4% (0.35 to 0.61 mM $Na^+$ per g of DW) (Volkov et al., 2003). Potassium contents dropped under the treatment for *Arabidopsis* both in roots and shoots, while on the opposite slightly increased in *Thellungiella* (Volkov et al., 2004). Kinetics of net sodium accumulation was registered over 72 hours and described by sum of two exponential summands with different coefficients; twice higher saturation values were observed for roots and shoots of glycophyte *Arabidopsis* compared to halophyte *Thellungiella* (Wang et al., 2006; Wang, 2006). Fast initial rates of net $Na^+$ uptake by roots during the first 6 h were 0.064 μmole/(g root DW*min) in *Thellungiella* and 0.048 μmole/(g root DW*min) in *Arabidopsis* (based on 0 and 6 hours data points). After 24 h of salt treatment, the rates dropped to 0.0004 nmole/(g root DW*min) in *Thellungiella*, and 0.003 nmole/(g root DW*min) in *Arabidopsis* (Wang, 2006).

However, to understand and potentially control ion transport the detailed mechanisms are needed, therefore it's important to use also the other methods. Another problem is to compare the ion fluxes measured by different methods and techniques, changes of ion concentrations are expressed in moles/g of fresh or dry weight (FW or DW), while electrophysiological and

MIFE measurements are normalised per a unit of surface area, $A/m^2$ and $moles/(m^2*s)$, correspondingly. Rough estimates could be helpful to translate different units to each other. Dry weight is about 10-15% of fresh weight, for recalculation per surface area a few assumptions and values concerning size of roots or protoplasts are required. Estimates relating net potassium flux to epidermal cells with sufficient and required ion currents to support the flux were done earlier for rye roots (White, Lemtiri-Chlieh, 1995). The net $K^+$ flux was estimated 1.0 - 1.9 µmole $g^{-1}$ root WF $h^{-1}$ and unidirectional $K^+$ was about 7 µmole $g^{-1}$ root FW $h^{-1}$ from mineral solution with 0.6 mM $K^+$ (White et al., 1991); epidermal cells with diameter 26 µm were considered 8.3% of root volume; then the net fluxes in µmole $g^{-1}$ root WF $h^{-1}$ corresponded to 0.11-0.21 pmole/(cell*hour) or 3.1-5.9 pA/cell. Unidirectional flux of 7 µmole $g^{-1}$ root FW $h^{-1}$ corresponded to 0.77 pmole/(cell*hour) or 21.7 pA/cell (White, Lemtiri-Chlieh, 1995). Converting to surface area of the whole protoplasts, it's possible to recalculate the fluxes being 0.11-0.21 pmole/(cell*hour) ≈ 14-27 nmole/($m^2$*s) and 0.77 pmole/(cell*hour) ≈ 100 nmole/($m^2$*s), the values reasonably confirmed by the other methods (see below). Conversion of the flux in pmoles to number of ions per a cell gives 0.77 pmole/(cell*hour) ≈ $1.3 * 10^8$ ions/(cell*second), coinciding with values and estimates at Figure 1 for the corresponding sizes of cells and ion currents via membrane.

Electrophysiological techniques of two-electrode voltage clamp or patch clamp measure ion currents across membrane of plant cells (protoplasts for patch clamp) under determined applied voltage via the membrane and provide the current-voltage curves, which are the functions of ion current (ion flux) depending on given voltage across the studied membrane (a sort of artificially applied membrane potential) (Figure 4). For the example of patch clamp study with root protoplasts of salt-sensitive *Arabidopsis* and salt-tolerant *Thellungiella* (Volkov, Amtmann, 2006) the inward ion fluxes of $K^+$ in external 100 mM KCl at -80 mV would correspond to 120 mA/$m^2$ =120 fA/$mkm^2$ ≈ $1.2 \cdot 10^{-18}$ mole/($mkm^2$*s) = 1.2 µmole/($m^2$*s) for *Arabidopsis* and 30 mA/$m^2$ ≈ 300 nmole/($m^2$*s) for *Thellungiella*, correspondingly. Inward sodium fluxes under the same conditions, but in 100 mM NaCl in the external medium for the protoplasts are 70 mA/$m^2$ ≈ 700 nmole/($m^2$*s) for *Arabidopsis* and 15 mA/$m^2$ ≈ 150 nmole/($m^2$*s) for *Thellungiella* (Volkov, Amtmann, 2006).

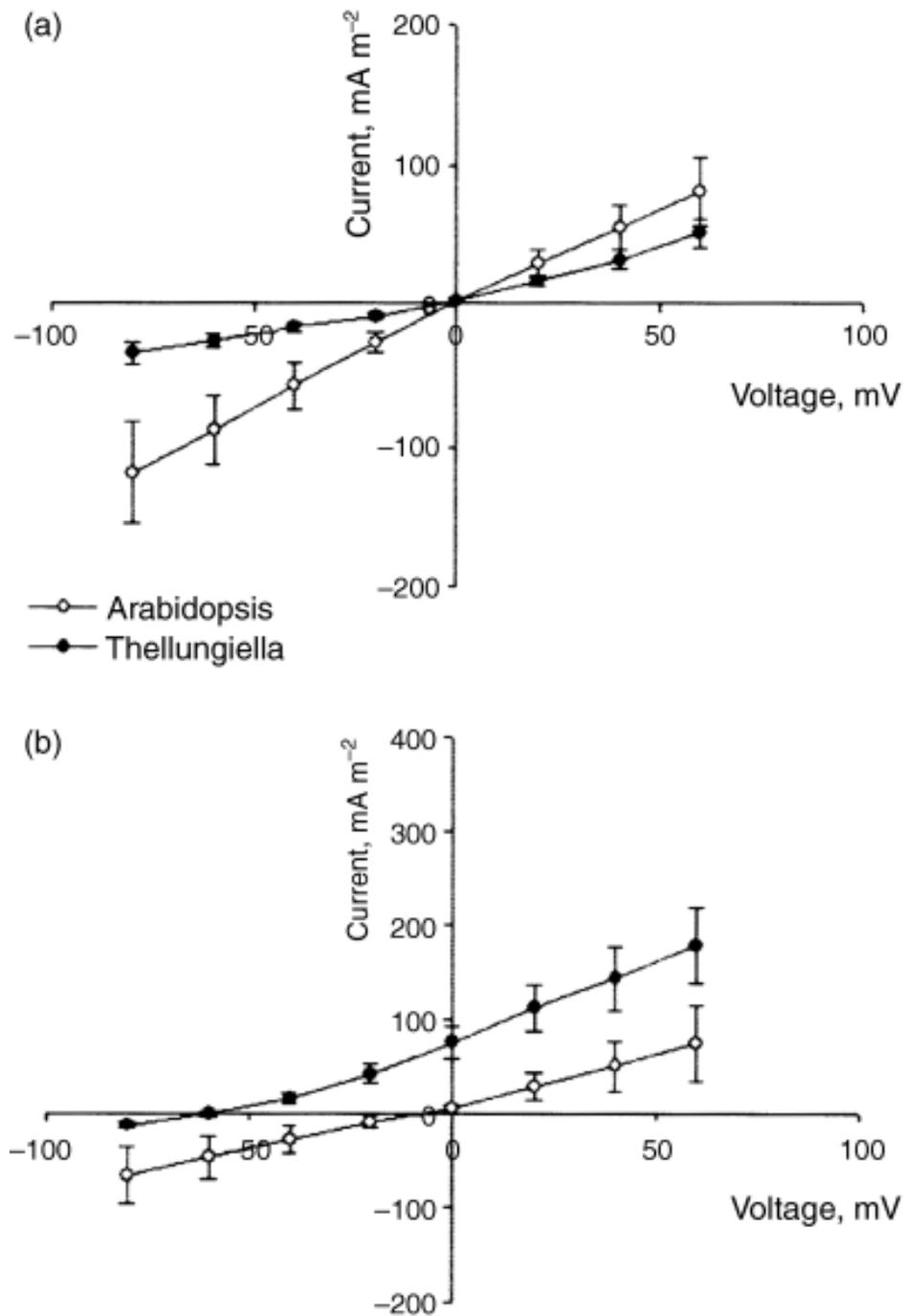

Figure 4. Comparison of current–voltage curves for whole-cell instantaneous currents in root protoplasts of *A. thaliana* (open symbols) and *T. halophila* (closed symbols). Currents are normalized to protoplast surface. The pipette solution was always 100 mM KCl. The bath solution was 100 KCl (a) or 100 NaCl (b). Data are given as means ± SE ($n = 6$ for *A. thaliana*, $n = 13$ for *T. halophila*). Reproduced from (Volkov, Amtmann, 2006) with the permission from the publisher John Wiley and Sons. Note that IV curve for instantaneous current in *T. halophila* resembles (though not completely obeys) the expected one from

Goldmann-Hodkin-Katz equation and the shift in reversal potential of ion current in 100 mM KCl/100 mM NaCl is reflected in slope of IV curve for voltages above and below the reversal potential.

Comparisons of electric currents and ion fluxes were performed in a series of simultaneous measurements using patch clamp and MIFE for wheat root protoplasts, when MIFE was used for measuring proton fluxes (Tyerman et al., 2001). Proton fluxes in the experiments basically correlated with ion currents, though not always with clear predicted coefficient of proportionality and with large variation between protoplasts (Tyerman et al., 2001). Further on MIFE proved the lack of "dark matter" and "dark" electroneutral fluxes for potassium transport in the wheat root protoplasts (the situation was different for $Ca^{2+}$, large $Ca^{2+}$ fluxes were not electrogenic) meaning that $K^+$ ion currents via potassium channels (outward KORC and inward KIRC) nearly exactly corresponded to the $K^+$ ion fluxes measured by ion selective electrodes of MIFE (Tyerman et al., 2001; Gilliham et al, 2006). Some deviations from 1:1 ratio and variation between protoplasts (ratio from 0.5 to 2.4) were explained by uneven distribution of ion channels in plasma membrane of protoplasts with clusters of ion channels; hence, fluxes measured by ion-selective microelectrode of MIFE system were smaller or larger depending on the position near the surface of protoplast (Gilliham et al, 2006). Potentially atomic force microscopy or ion-conductance scanning microscope with further recording of ion channel activity (e.g. Hansma et al., 1989; Korchev et al., 1997; Lab et al., 2013 for ion-conductance scanning microscope) could be useful to explore the distribution of lipid rafts and clusters of ion channels in the plasma membrane of protoplasts.

Usually reported patch clamp values for ion fluxes in protoplasts are higher than fluxes measured by vibrating ion-selective electrodes (MIFE) for intact roots of several plant species (e.g.: Shabala, Lew, 2002; Chen et al., 2013), but mostly coincide within orders of magnitude and essentially depend on composition of ambient medium, concentration of ions, membrane potential of cells and on multiple physiological factors. For example, the above mentioned inward $K^+$ fluxes measured by patch clamp in external 100 mM KCl at -80 mV correspond to 1200 nmole/($m^2$*s) for *Arabidopsis* root protoplasts (Volkov, Amtmann, 2006), the patch clamp $K^+$ fluxes could be over ten times higher at more negative voltages with active potassium inward rectifying channels (e.g. Ivashilina et al., 2001), while net potassium influx about 2-3 nmole/(m$^2$*s) was recorded using MIFE in much lower 1 mM KCl from roots of maize under darkness and –4-5 nmole/(m$^2$*s) (negative flux, efflux) after shoot illumination (Shabala et al., 2009). However, the fluxes are strongly influenced and for barley root segments in 0.5 mM KCl potassium flux of about -100 nmole/(m$^2$*s) (efflux) was reverted to over +250 nmole/(m$^2$*s) (influx) by application of 4 μM of cytokinin-like phytohormone kinetin (Shabala et al., 2009).

Experiments with MIFE are non-invasive and simpler, so provide huge opportunities with temporal resolution of seconds and spatial resolution within less than tens of microns for exploring physiological factors and conditions, which influence ion fluxes to and out of cells (Newman et al., 1987; Newman, 2001; Shabala, 2006; Sun et al., 2009; Shabala, Bose, 2013). Ion fluxes of potassium, sodium and the other ions in the vicinity of roots were measured by MIFE and compared for different salt-tolerant and salt-sensitive cultivars and agricultural species (Chen et al., 2005; Cuin et al., 2008; Cuin et al., 2012), along root zones of several plants (Garnett et al., 2001; Chen et al., 2005; Pang et al., 2006), for mutants in specific ion channels or transporters (e.g. Shabala et al., 2005; Demidchik et al., 2010), under treatment by physiologically active compounds and molecules (Cuin, Shabala, 2007; Shabala et al., 2009; Pandolfi et al., 2010; Demidchik et al., 2011; Ordoñez et al ., 2014), after generation of reactive oxygen species and salt stress (Cuin, Shabala, 2007b; Demidchik et al., 2010) and for numerous other conditions. Due to high resolution and being non-invasive MIFE also could be a very good method for studying ion transport in cell biology (Lew et al., 2006; Valencia-Cruz et al., 2009; Demidchik et al., 2010). The results provided huge volume of information about characteristics, kinetics and physiological features of ion fluxes under salt stress, helped to develop fast tests for salinity tolerance (Chen et al., 2005; Cuin et al., 2008). The results are described in several reviews (e.g. Newman, 2001; Shabala, 2006; Sun et al., 2009; Shabala, Bose, 2013) and hundreds of publications, so will not be covered here in more detail. Among the limitations of MIFE is the selectivity of ion-selective electrodes, which is influenced by interfering ions (e.g. Knowles, Shabala, 2004) and sometimes affected by physiologically active compounds and proteins in the medium for measurements after interaction with the material of ion-selective electrodes (Volkov, de Boer, unpublished results; Chen et al., 2005), so again more control checks are required.

Unidirectional fluxes of $^{22}$Na$^+$ for sodium and $^{42}$K$^+$ or Rb$^+$ being a relative tracer for potassium are helpful for fast kinetics of ion transport within tens of seconds determined by the speed of sampling and changes in concentrations at the background of initial level lacking

$^{22}$Na$^+$ and $^{42}$K$^+$/Rb$^+$ (e.g. Figure 5 for $^{22}$Na$^+$ fluxes). The method is used widely and had already provided essential advances in plant ion transport (e.g. MacRobbie, Dainty, 1958; Rains, Epstein, 1965; Epstein, 1966). The outward fluxes could be measured after loading plants with $^{22}$Na$^+$ or $^{42}$K$^+$/Rb$^+$ and transferring then to different chemical solutions without the ions (e.g. Wang et al., 2006; Wang, 2006 for $^{22}$Na$^+$ for roots of *Arabidopsis* and *Thellungiella*). Details and the methodical procedures are well described with different modifications to get more information about compartmentation of the absorbed ions and to exclude potential sources of errors, which may have an effect on the results and their interpretation (Cheeseman, 1986 for analysis of compartmentation based on efflux kinetics; Wang, 2006; Britto, Szczerba, Kronzucker, 2006 for $^{42}$K$^+$ in application for the tracer efflux by barley roots; Britto, Kronzucker, 2013 for comprehensive practical description of experimental procedures to measure potassium fluxes). The values of measured fluxes differ significantly, sometimes over 100 times depending on plant species, physiological conditions and ion concentrations (e.g. summarised data for sodium fluxes in: Kronzucker, Britto, 2011).

Analysis of unidirectional fluxes is often complemented by the other methods to obtain better understanding of the processes and the results. An interesting comparison for influx of K$^+$ ($^{86}$Rb$^+$) from a 0.1 mM K$_2$SO$_4$ solution and net K$^+$ influx obtained using external K$^+$ microelectrodes (prototype of MIFE) gave nearly the same values of fluxes: 2.6 μmole/(g FW*h) and 2.5 μmole/(g FW*h), correspondingly (Newman et al., 1987). Measurements of unidirectional $^{24}$Na$^+$, $^{42}$K$^+$ fluxes in barley supported by membrane potential measurements and pharmacological profiling of fluxes allowed to study high affinity transport of sodium from low μM – 50 mM solutions and provided predictions about possible mechanisms for the transport (Schulze et al., 2012).

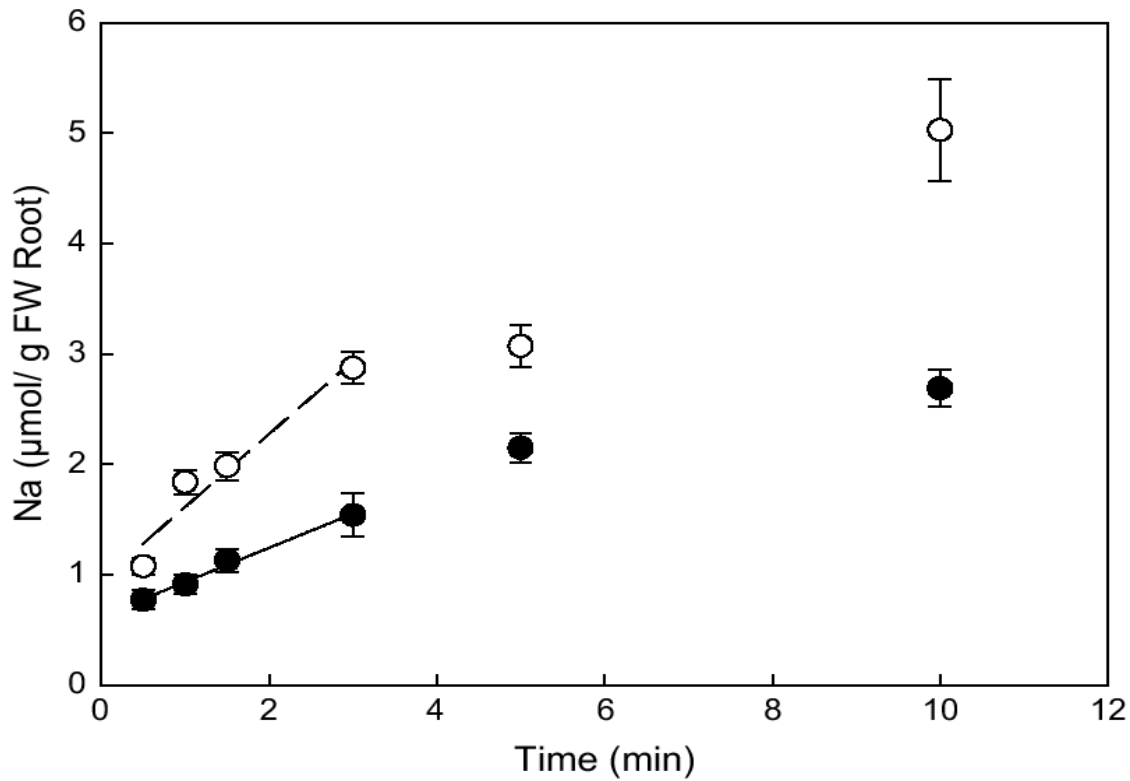

Figure 5. Kinetics of initial unidirectional Na$^+$ influx into roots of *A. thaliana* (open circles) and *T. halophila* (closed circles) as determined from $^{22}$Na$^+$ accumulation of individual plants from $^{22}$Na$^+$ labelled nutrient solution with 100 mM NaCl and 0.1 mM CaCl$_2$. Error bars are SE (n=4). Reproduced from (Wang et al., 2006) with the permission from the publisher Oxford University Press.

Among methods to determine ion fluxes is the use of ion-selective fluorescent indicators for estimating cytoplasmic and vacuolar concentrations of sodium and potassium together with kinetics of their changes (e.g. mostly for protoplasts: Lindberg, 1995; Halperin, Lynch, 2003; D'Onofrio, Kader, Lindberg, 2005; Kader, Lindberg, 2005), measurements by intracellular ion-selective electrodes (e.g. Carden et al., 2003 and references there), $^{23}$Na-NMR spectroscopy (e.g. Bental, Degani, Avron, 1988) and several others; the methods are not discussed in detail here. Independent combination of several approaches and methods with further comparisons and recalculations of fluxes is essential to get reliable results and conclusions.

# Ion transport in halophytes and effects of salt stress on membrane transport in glycophytes

Surprising examples of salinity tolerance are represented by halophytes, which are able to grow at high concentrations of salt (Figure 6), under irrigation by seawater and even under several times higher salt concentrations than in seawater (over 2 M of NaCl) (Flowers, Troke, Yeo, 1977; Flowers, Colmer, 2008). Halophytes is an undisputable example and proof that salinity tolerance in plants in achievable. The main question is how to bring the trait or multiple traits to agricultural plants and to which extent it could be realised without essentially influencing the growth rate, agricultural productivity and quality of grain, fruits, edible parts, flowers or the other economically important elements/features/traits of agricultural plants.

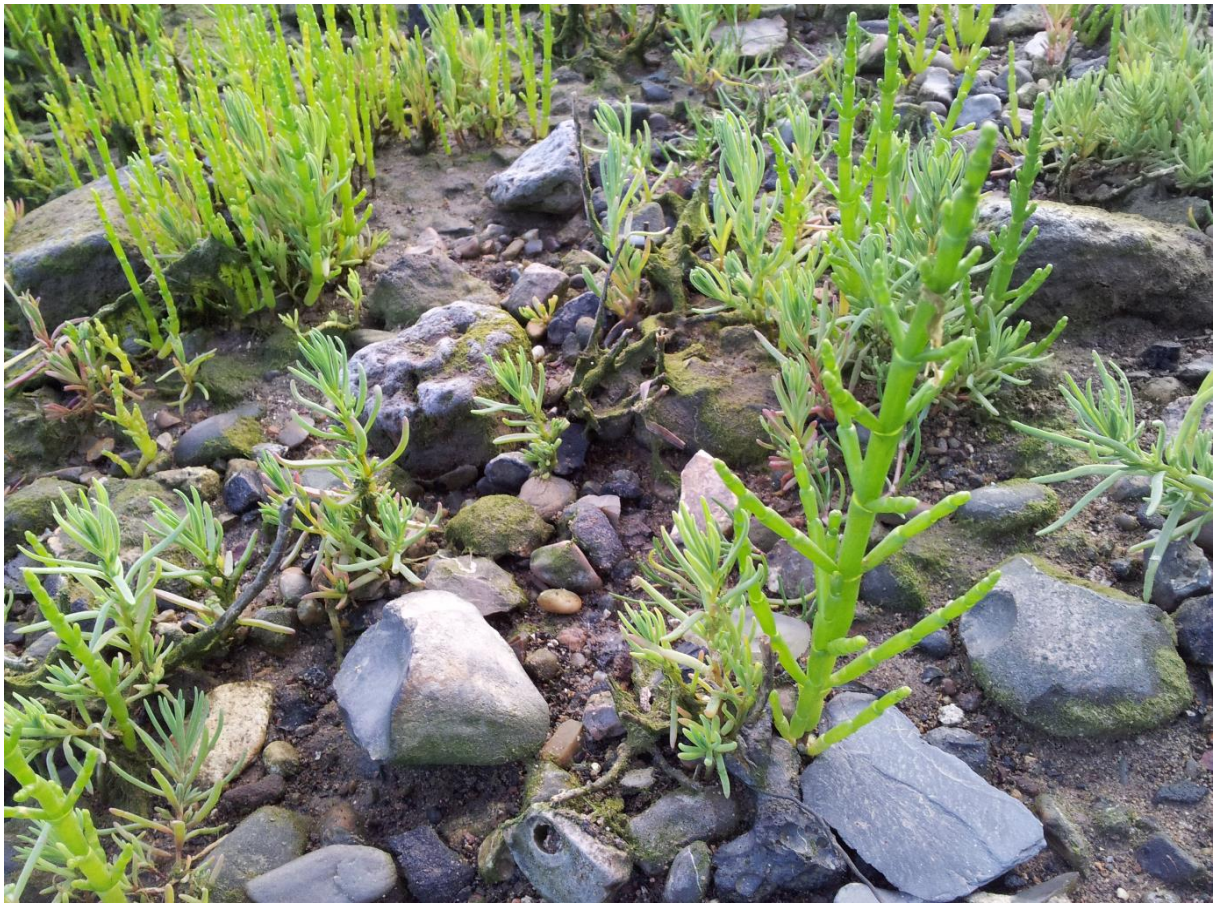

Figure 6. Halophytes *Salicornia sp.* and *Suaeda maritima* are growing at the salt-affected soil near river Medway in the UK, where the area is flooded with mixed sea and river water under high tides. Beginning of June 2014. The size of the largest specimen is about 15 cm.

Halophytes are usually considered like sodium tolerant plants since NaCl is the main source of salinity in most areas, though chloride, sulphate, calcium, magnesium and the other ions are involved constituting sometimes the main salt for soil and water salinisation (Flowers, Troke, Yeo, 1977). Halophytes imply several strategies to cope with high concentrations of salts including sodium exclusion from roots, accumulation of high sodium in shoots, shedding specialised leaves, localising salts in vacuoles, excreting them via salt glands etc.; the role and contribution of each strategy depends on the habitat and type of a halophyte plant (Flowers, Troke, Yeo, 1977; Breckle, 2002; Lüttge, 2002; Flowers, Colmer, 2008; Munns, Tester, 2008; Shabala, 2013). So far the known transport systems in halophyte plants are basically the same like in glycophytes due to common ancestry and evolution (Flowers, Galal, Bromham, 2010), when, for example, the trait of salt tolerance emerged independently over 70 times in different groups of grasses (Bennett, Flowers, Bromham, 2013). The knowledge facilitates potential transfer of salinity tolerance traits linked to ion transport to agriculturally important plants. "Domestication" of halophyte plants is the other way to use salt-affected and salinized territories.

It should be, however, emphasised that the simple idea of straightforward and strict decrease in sodium net influx and increasing potassium net uptake via plasma membrane of epidermal root protoplasts to alter mineral nutrition under salinity may not and the most likely will not be productive (Flowers, 2004; Maathuis, Ahmad, Patishtan, 2014). From one point, salt tolerance could be linked with the other tissues, vacuolar accumulation of sodium or/and e.g. xylem/phloem transport properties. From another point, a living cell is a complex system. Increase in cytoplasmic sodium and reduction of $K^+$ result in changes of membrane potential, osmotic pressure, turgor pressure, calcium signalling, reactive oxygen species signalling, transcriptional regulation, alteration of gene expression, modification of protein expression pattern and spectra of siRNAs, signalling molecules and metabolites. The complex network of signalling events and metabolic reactions is involved and based on the new dynamic attractor including the ion fluxes. The volume and shape of the attractor in multidimensional space could be registered using means of "omics" that is RNA expression microarrays, proteomics, ionomics, metabolomics etc. (Figure 7). The properties of the attractors are slowly studied and understood for the biological systems (Spiller et al., 2010; Breeze et al., 2011), though the idea is well developed in physics and especially in plasma physics for non-equilibrium thermodynamic systems (Akhromeeva et al., 1989; Akhromeeva et al., 1992). Comparison of ion fluxes via membranes between halophytes and glycophytes often demonstrates lower sodium uptake fluxes for halophytes (reviewed in: Flowers, Colmer, 2008). However, an evident problem in comparison between halophytes and glycophytes is the high variability in ion transport between plant species linked with e.g. growth rate, tissue-specific variability and the other physiological factors. It's a need to consider similar plants and achieve comparable values. Recent results on ion fluxes in glycophyte *Arabidopsis* and close from the point of genome and morphology halophyte *Thellungiella* demonstrated lower $Na^+$ fluxes and higher $K^+/Na^+$ selectivity of ion currents in the roots and root protoplasts of the halophyte under salt treatment (Figures 4, 5 and in: Volkov et al., 2004; Wang et al., 2006; Wang, 2006; Volkov, Amtmann, 2006; similar results for roots of the two plants: Alemán et al., 2009). However, different strategies may be involved depending on the level of salinity tolerance, plant morphology, habitat and the other environmental factors and evolutionary history. For example, salt tolerant *Plantago maritima* had similar sodium uptake rates by roots compared to salt-sensitive *Plantago media* (Erdei, Kuiper, 1979; de Boer, 1985); the salt-tolerance in the pair is rather associated with xylem transport and sodium accumulation in vacuoles of leaf cells.

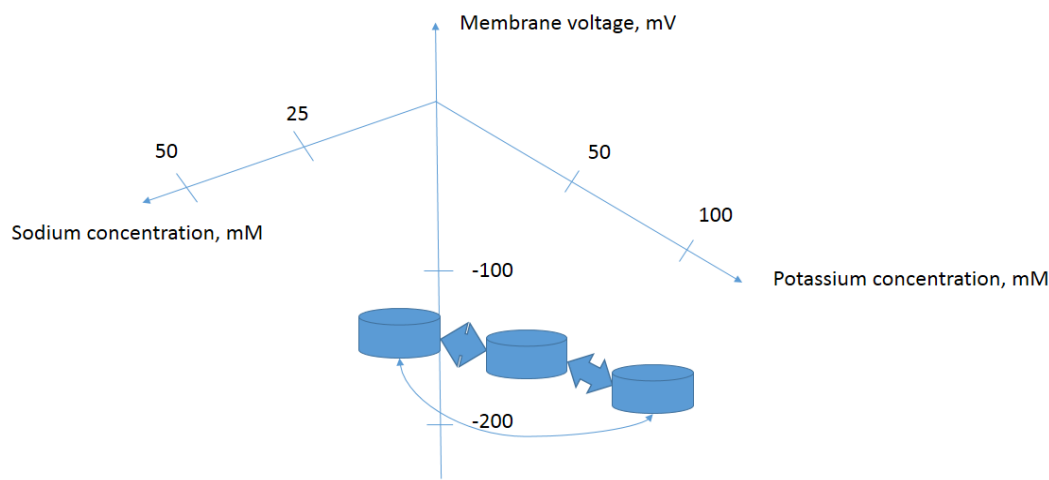

Figure 7. Proposed simplified model of several attractors for a plant cell with specific metabolic and regulatory networks determined by cytoplasmic ion concentrations and membrane potential. Changes in extracellular ion concentrations result in transition of cells from one state of concentrations-membrane potential-protein and DNA-RNA expression levels and activity pattern to another one. Stability of the biophysiological states and trajectories of transitions could be studies in more detail using complex approaches of ionomics-metabolomics-proteomics-nucleic acids expression arrays apart from biophysical methods.

Vacuolar membranes of halophytes were also a subject of special investigation. Patch clamp study of vacuoles from leaves of *Suaeda maritima* did not find any unusual features supporting high salt tolerance of the halophyte (Maathuis, Flowers, Yeo, 1992). Patch clamp experiments to compare vacuoles from roots of *Plantago maritima* and *Plantago media* also did not reveal striking differences apart from an extra smaller ion channel conductance in the tonoplast of the halophyte; salt stress essentially reduced the open probability of larger nonselective between $K^+$ and $Na^+$ ion channel conductance in both species without changing the properties of the conductance (Maathuis, Prins, 1990). However, comparison of tonoplast from suspension culture cells of halophytic sugar beet with glycophytic tomato revealed rectification properties of ion channels in vacuolar membrane of the halophyte (Pantoja, Dainty, Blumwald, 1989). At positive voltages in outside-out configuration corresponding to ion currents out of vacuole the ion-channel-like conductance dropped 6.5 times presumably preventing the transport of ions from vacuole to cytoplasm; the conductance was not selective for $K^+$ over $Na^+$ in both species (Pantoja, Dainty, Blumwald, 1989).

The complexity of ion transport and its regulation within the whole plant and specific changes in sodium and potassium fluxes under salinity are confirmed in experiments with glycophytes. It's reasonable to mention that ion transport may be essentially influenced by salt stress in glycophytes in a cell-specific manner and the knowledge is required to consider the whole plant responses to salinity and to alter them in a desired direction. An example of changes in membrane conductance after salt stress in several types of cells from barley leaves is given in Figure 8.

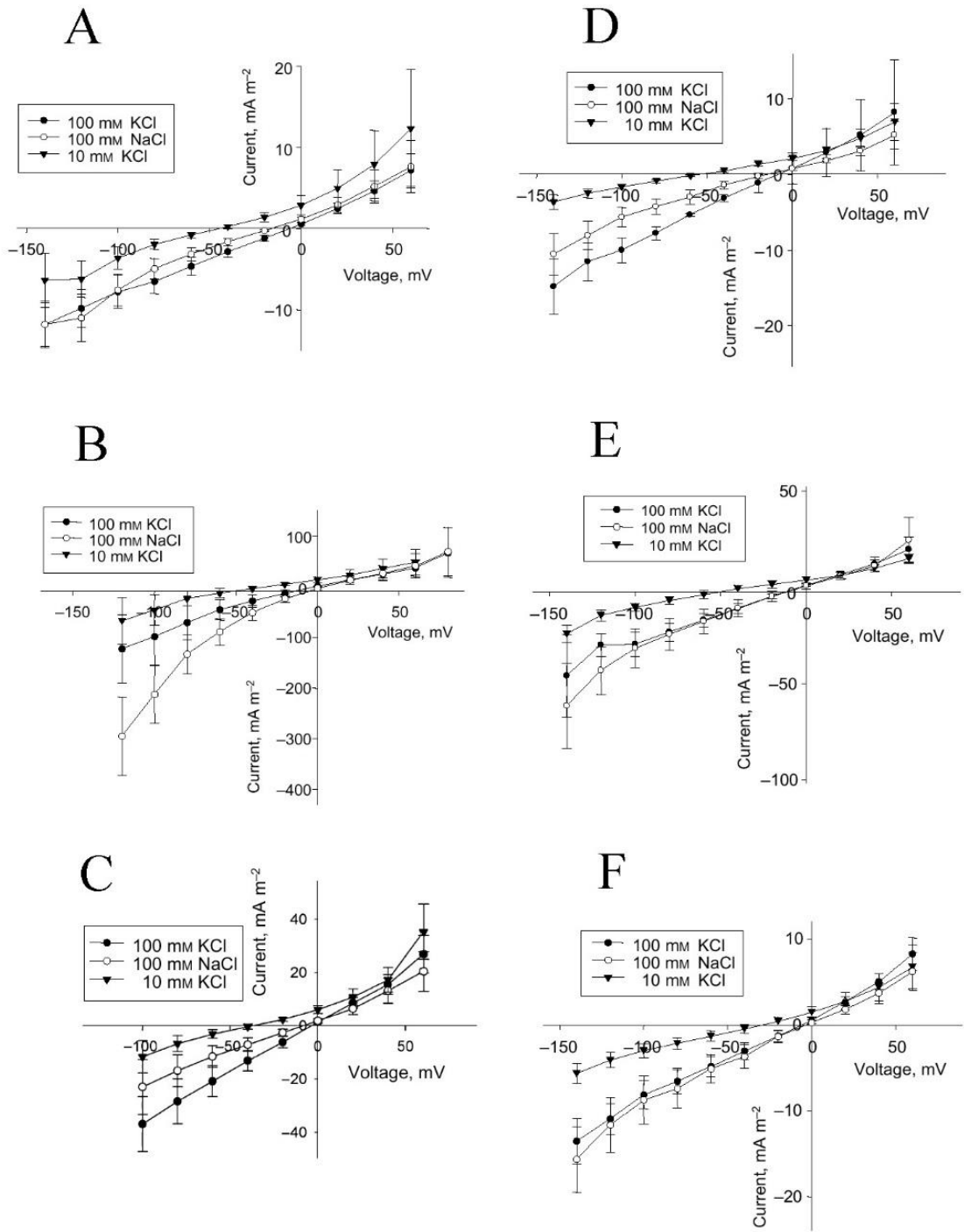

Figure 8. Effect of salt stress on instantaneous current in protoplasts from the elongation zone and emerged blade portion of the developing leaf 3 of barley *(Hordeum vulgare* L.) (A–F). Averaged current–voltage *(*I–V) relationship in protoplasts from mesophyll (A, D) and epidermis (B, E) of the emerged blade, and from protoplasts of the elongation zone (C, F).

Control (A-C) and salt treatment (D-F), $n$ =5-8 for control protoplasts and $n$ =3–6 for protoplasts for salt treatment; error bars are standard errors. Concentrations of KCl and NaCl in the bath are indicated. The pipette solution was always 100 mM KCl. Plants had been exposed to 100 mM NaCl for 3 days prior to protoplast isolation. Reproduced from (Volkov et al., 2009; composed from Figures 1, 2, 4, 5) with the permission from the publisher John Wiley and Sons.

## Genetic engineering with tissue-specific overexpression/knockout/disruption of specific transporters modifies salinity tolerance

Essential rise and success of molecular methods together with the identification and characterisation of individual ion channels and transporters allowed to understand their role in sodium and potassium transport under salinity and their importance for salinity tolerance. Several successful attempts to overexpress or knockout genes of HKT, NHX or SOS1-like transporters and vacuolar proton pump $H^+$-PPase to alter the salinity tolerance of plants had been reported.

Overexpression of vacuolar $H^+$-PPase sharply increased salinity tolerance in *Arabidopsis* (Gaxiola et al., 2001). Overexpressing plants accumulated more sodium and potassium in their leaves and also demonstrated higher drought resistance. The overexpression of the vacuolar $H^+$-pump enhanced the proton pumping activity at vacuolar membrane and thus permitted to accumulate more $Na^+$ in vacuoles due to activity of $Na^+/H^+$ antiporters NHX. The choice of $H^+$-pyrophosphatase is explained by a single gene required for the protein, while the other vacuolar $H^+$-ATPase is composed of several subunits and needs correct overexpression of several genes. Further attempts to overexpress vacuolar $H^+$-PPases from different plant and microbial species increased salinity tolerance in tobacco (D'yakova et al., 2006; Gao et al., 2006), transgenic rice overexpressing NHX1 (Zhao et al., 2006), alfalfa (Bao et al., 2009), cotton (Pasapula et al., 2011) and tomato (Bhaskaran, Savithramma, 2011). Gene of vacuolar $H^+$-pyrophosphatase was found to be among salinity tolerance determinants in barley (Shavrukov et al., 2013).

Another vacuolar protein is Na$^+$/H$^+$ antiporter NHX, which is important for sequestration of Na$^+$ in vacuoles at the expenses of H$^+$ gradients created by vacuolar H$^+$-ATPases pumping protons inside the vacuole. Overexpression of NHX increased salinity tolerance in *Arabidopsis*, the overexpressing plants accumulated more Na$^+$ compared to wild type and demonstrated higher Na$^+$/H$^+$ exchange activity in isolated leaf vacuoles (Apse et al., 1999). The approach of overexpressing NHX1 from *Arabidopsis* to improve salinity tolerance proved to be successful for tomato, the transgenic plants accumulated more sodium in leaves but not in fruits (Zhang, Blumwald, 2001). Cotton plants with AtNHX1 from *Arabidopsis* (He et al., 2005), rice overexpressing SsNHX1 from halophyte *Suaeda salsa* (Zhao et al., 2006), tomato with heterologous Nhx from *Pennisetum glaucum* (Bhaskaran, Savithramma, 2011) also showed increased salinity tolerance. The results with heterologous expression or overexpression of NHX transporters lead to conclusions that the gene is among determinants and potential candidates for engineering salinity tolerance (e.g. Peleg, Apse, Blumwald, 2011 with more references for successful overexpression of NHX to increase salinity tolerance in sugar beet, wheat, maize and the other plants).

The amazing simplicity of the idea to play with the expression of known and functionally well characterised transporters and get a salt tolerant or salt sensitive plant is applied to plasma membrane SOS1 Na$^+$/H$^+$ antiporters and Na$^+$ or Na$^+$/K$^+$ HKT transporters. *Arabidopsis* mutants with defects in gene of SOS1 transporter exhibited strong growth inhibition under salt treatment (Wu, Ding, Zhu, 1996), which was rescued in *sos1* mutant by overexpression of SOS1 gene under strong 35S viral promoter (Shi et al., 2000). Overexpression of SOS1 gene in wild type plants enhanced salinity tolerance of *Arabidopsis*, reduced sodium accumulation in shoots and sodium concentration in xylem sap (Shi et al., 2003). Further on overexpression of SOS1 from *Arabidopsis thaliana* increased salinity tolerance in transgenic tobacco (Yue et al., 2012) and in transgenic tall fescue (Ma et al., 2014); SOS1 gene from durum wheat conferred salinity tolerance to *sos1* mutant of *Arabidopsis* (Feki et al., 2014). Disruption of SOS1 activity by RNA interference in *Thellungiella* on the opposite resulted in the loss of tolerance of the halophyte indicating importance of Na$^+$ efflux and essential role of SOS1 in salinity tolerance (Oh et al., 2009a). It is interesting to mention that RNA interference of SOS1 significantly changed the whole transcriptome of *Thellungiella* (Oh et al., 2007) and vacuolar pH under salt treatment (Oh et al., 2009b) indicating the complex nature of metabolic and regulatory networks in plants (Figure 7) and the probabilistic chances of success in strict overexpression of specific transporters for salinity tolerance improvement. It is also worth to mention that overexpression of SOS1 and its regulatory components together with NHX1 in *Arabidopsis* to increase the tolerance more unexpectedly did not confirm the earlier results with NHX1 (Yang et al., 2009) and need further detailed investigation. A more complicated situation emerges due to tissue-specific expression. SOS1 seems to be important for long-distance ion transport and xylem loading/unloading in *Arabidopsis* (Shi et al., 2002), sodium partioning between plant organs in tomato (Olías et al., 2009) and ion fluxes in root meristem zone (Guo, Babourina, Rengel, 2009) being expressed in plasma membrane of xylem parenchyma and root apical epidermal cells (Shi et al., 2002).

Genetic modification of salinity tolerance using HKT transporters was also successful. Analysis of *Arabidopsis* plants with mutated HKT gene revealed higher salt sensitivity of the mutants under long term stress, higher sodium accumulation in their shoots under mild salinity treatment (Mäser et al., 2002) and suggested that HKT is involved in recirculation of sodium within plants (Berthomieu et al., 2003). Further study confirmed increased sodium in the shoots of *Arabidopsis hkt1;1* mutants and clarified that HKT is important for root accumulation of $Na^+$ and $Na^+$ uptake from xylem (Davenport et al., 2007). The next step was to create plants overexpressing HKT (Møller et al., 2009). *Arabidopsis* plants overexpressing AtHKT under the control of strong 35S promoter were compared with plants specifically overexpressing HKT in cells of root stele; Pro35S:HKT1;1 plants were salt sensitive probably due to higher $Na^+$ uptake by roots while tissue specific overexpression of HKT in stele increased salinity tolerance and reduced sodium accumulation in shoots (Møller et al., 2009). The approach was applied to rice where gene from *Arabidopsis* AtHKT1;1 was heterologously expressed in root cortex, it resulted in lower shoot $Na^+$ concentrations and improved salinity tolerance (Plett et al., 2010). HKT transporters proved to be important for $Na^+$ exclusion in wheat and were transferred from durum wheat to bread wheat by interspecific crossing; the genes gave beneficial effects including higher $K^+/Na^+$ ratio in leaves under saline conditions (James et al., 2011). The results set HKT transporters to potential candidates for engineering salinity tolerance and among the determinants of the trait (reviewed in: Horie, Hauser, Schroeder, 2009; Almeida, Katschnig, de Boer, 2013) together with the above mentioned NHX1, SOS1 and presumably new studied transporters, e.g. similar to CHX21 from *Arabidopsis* (Hall et al., 2006).

# Perspectives of protein engineering. Structure-function studies and potential future for expression of novel ion channels and transporters

Novel opportunities for increasing salinity tolerance in plants are arising with the progress of new methods of molecular biology, understanding regulation networks from synthetic biology and growing knowledge about single point mutations changing specific amino acids within molecules of an ion channel or a transporter.

Single amino acid substitutions, e.g. within $K^+$ selectivity filter GYG of potassium channels, may change selectivity of the ion channels rendering them from $K^+$ selective to nonselective (eg. Heginbotham, Abramson, MacKinnon, 1992). Amino acid substitutions within a presumed modelled pore region of HKT transporters are able to alter them from $Na^+$ selective to $Na^+$ and $K^+$ permeable (e.g. Mäser et al., 2002; Almeida, 2014); moreover the single point mutations could be determining for salinity tolerance, e.g. amino acid substitution V395L in rice transporter OsHKT1;5 presumably explained the salt sensitivity of rice cultivar (Cotsaftis et al., 2012). Specific single point substitution in *Arabidopsis* HAK5 transporter (F130S) over 100 times increased affinity for $K^+$ under heterologous expression and reduced inhibition constants for $Na^+$ and $Cs^+$ (Aleman et al., 2014). Effects of single point mutations on the whole pattern of physiology and on phenotype are well known from human biomedical science when inherited diseases cystic fibrosis and sickle cell anemia are caused by amino acid substitutions in transport protein CFTR and in haemoglobin, correspondingly.

The knowledge about structure-function correlations of proteins allows to modify the selectivity and create the required properties of ion transport proteins. The way in the direction is to employ the existing and growing information about structure-function of different ion channels and transporters, including plant (with special attention to halophytes), animal, algal, bacterial and fungal ones, change their ion selectivity and gating properties according to the requirements using single amino acids mutations and transform the plants of interest in a tissue-specific or cell-specific manner. Examples of tissue-specific transformation already exist (e.g. Møller et al., 2009; Plett et al., 2010) and the new methods and opportunities are progressing enormously (e.g. Brandt et al., 2006; Oszvald et al., 2008 etc.).

An alternative approach from synthetic biology is not to modify the existing membrane transport proteins, but to create new ones with desired properties for the further cell-specific transformation (Figure 9). The approach had not been widely used and may bring fruitful results. The idea is different from what could be assumed at a first glance. The existing biological organisms emerged over the process of long evolution, when previous "building blocks" and elements were used for the future development and often could not be essentially modified due to intrinsic links within organisms and biological systems. It leaves out the question of ideal design, which is mostly not present in biological organisms since they are largely predetermined by the previous evolutionary history with intrinsic evolutionary trajectories and evolved under multifactor environment (composition of atmosphere, illumination, temperature, salinity and mineral nutrients, water availability etc., while interactions with the other organisms and biotic interaction being the most important for most situations). A simple example is related to temperature. Ion channels in homeothermic animals like mammals or birds evolved over hundreds of millions of years under stable conditions and hence differ in many properties from ion channels in plants with specific regulation networks. Sodium and calcium selective ion channels had not been found in plants while in animals they ensure action potentials in neurons and cardiomyocytes and their functioning. Specialised highly temperature-sensitive ion channel in animals provide temperature sensation (for example transient receptor potential channels, e.g.: Ramsay, Delling, Clapham, 2006; Myers, Sigal, Julius, 2009). Plants are different, they rely on calcium signalling via nonselective cation channels, may have distinct groups of transporters with specific properties, have no known sodium-selective ion channels and use ion channels with low temperature sensitivity (e.g. plant potassium channels and their regulation are reviewed in: Dreyer, Uozumi, 2011). Obviously, cell signalling and regulation in plants have numerous specific peculiarities compared to animals.

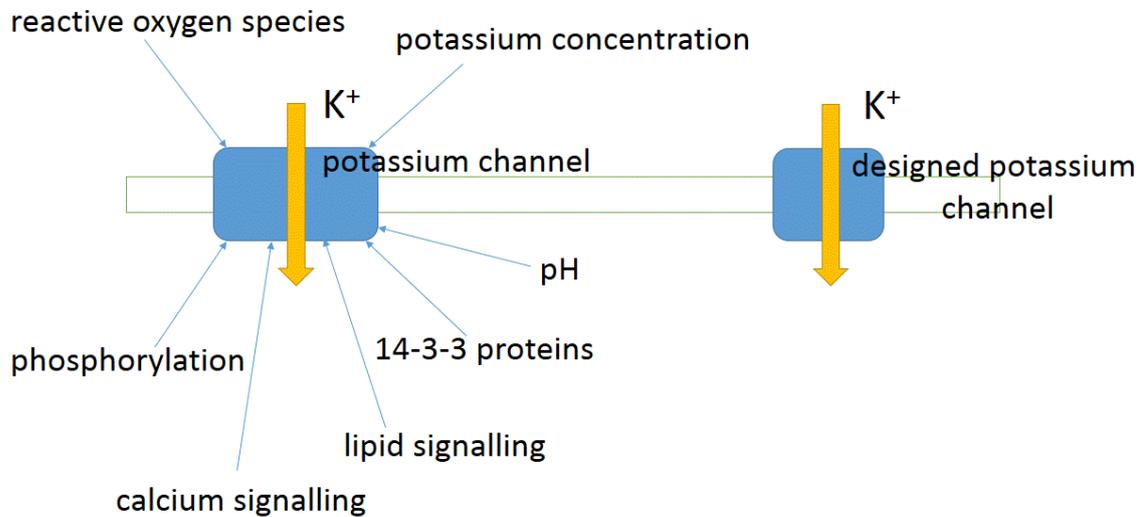

Figure 9. Novel artificially designed ion channels and ion transporters potentially provide an opportunity to escape the present evolved protein regulation networks and may be useful for altering ion concentrations, membrane potential and signalling without essentially interacting with the fine-tuning of the existing regulatory networks. A schematic generalised plant potassium channel in a plasma membrane with several potentially known regulation factors and a model of artificially designed ion channel lacking the regulation feedbacks.

From the point of the above mentioned it seems that attempts to design novel artificial ion channels and transporters with known characterised ion selectivity filters and voltage sensors adding or excluding specific interacting regulatory elements for the proteins might be productive. The ion channels and transporters when expressed in cell-specific manner under controlled conditions and in defined numbers may avoid fine tuning of regulation and potentially could provide shortcuts in natural signalling networks. The appearing opportunities offer new chances to design salt tolerant plants with previously unknown features and wider ranges of regulation circuits and networks.